\documentclass[reprint,amsmath,amssymb,aps,prl,longbibliography]{revtex4-1}

\usepackage{graphicx}
\usepackage{dcolumn} 
\usepackage{bm} 
\usepackage{physics}
\newcommand{\av}[1]{\langle {#1} \rangle} 

\begin{document}

\preprint{APS/123-QED}

\title{Out-of-equilibrium fluxes shape the self-organization of locally-interacting turbulence }

\author{Anton Svirsky$^{1}$ and Anna Frishman$^1$}
\email{frishman@technion.ac.il}

\affiliation{$^1$Physics Department, Technion Israel Institute of Technology, 32000 Haifa, Israel}

\date{\today}

\begin{abstract}
    We study the self-organization of turbulence in a geophysically motivated two-dimensional fluid with local interactions. Using simulations and theory, we show that the out-of-equilibrium flux to small scales imposes a constraint on the large-scale emergent flow. Consequently, a rich phase diagram of large-scale configurations emerges, replacing the unique state found in flows with energy injection below the interaction scale. We explain what sets the boundaries between the different phases, and the occurrence of spontaneous symmetry breaking. 
    Our work demonstrates that the selection mechanism of large-scale structures in quasi-geostrophic flows can be dramatically altered by forcing above the interaction scale.

\end{abstract}

\maketitle

Turbulent flows are shaped by conserved quantities. This is most evident in two-dimensional flows, where the coexistence of two conserved quantities leads to an inverse flux of energy: from small to large scales. 
In a finite domain, energy accumulates on large scales and a strong mean flow, called a condensate, spontaneously forms ~\cite{kraichnan_inertial_1967}. This phenomenology minimally captures the dynamics of flows constrained by rotation (small Rossby number $\rm{Ro}\ll1$) and stratification (small Froude number $\rm{Fr}\ll1$), as astrophysical and geophysical flows often are. Indeed, a cornerstone of the theory of large-scale geophysical dynamics, including the formation of coherent jets and vortices, is the (two-dimensional) quasi-geostrophic (QG) framework~\cite{vallis_atmospheric_2017}. In QG, there is a typical scale controlling the range of fluid-element interactions, called the Rossby deformation radius $L_d$.  Two-dimensional Navier-Stokes (2DNS) is recovered when $L_d\to \infty$, and condensates have been studied in detail in this limit~\cite{bose_smith_1993,dynamics_chertkov_2007,Universal_chertkov_2010,laurie_universal_2014,kolokolov_structure_2016,frishman_turbulence_2018,frishman_culmination_2017}, including in the presence of differential rotation (beta effect), e.g.~\cite{Farrell2007,Srinivasan2012,Tobias2013}. However, a fundamental understanding of whether and how condensates are influenced by a finite range of interactions is currently lacking. 
     
As the condensate is a self-organized structure, a basic question is what determines its form. When the forcing scale is very much smaller than that of the domain, the condensate is expected to be universal~\cite{laurie_universal_2014}, satisfying two guiding principles: (i) taking the largest available scale and (ii) conforming with the symmetries of the domain. These principles are born out in simulations of 2DNS~\cite{bouchet_random_2009,frishman_jets_2017,guervilly_jets_2017,Julien_impact_2018,xu_fluctuation-induced_2023}. As these principles do not invoke the out-of-equilibrium fluxes sustaining the system, they motivate characterizing the condensate using equilibrium statistical mechanics, e.g.~\cite{bouchet_random_2009,bouchet_statistical_2012,Gallet_two_2024}, theories originally developed for inviscid or decaying turbulence~\cite{miller_statistical_1990, sommeria_final_1991,montgomery_relaxation_1992,carnevale_rates_1992}.    

 Here we show how these expectations are broken when energy is injected at a scale larger than the interaction scale. We consider the extreme case of zero-range interactions, $L_d\to 0$, termed the large-scale quasi-geostrophic (LQG) model, expected to be relevant in the regime $\rm{Fr}^2\ll\rm{Ro}\ll \rm{Fr}\ll1$ ~\cite{polvani_coherent_1994,svirsky_two-dimensional_2023}. We find that, unlike 2DNS in a square periodic domain, the emergent condensate can take a multiplicity of configurations, including jets which spontaneously break the domain symmetry. Thus, while principle (i) still holds, the second principle can be broken.
We identify the two control parameters for the mean flow phase diagram and chart-out the corresponding bifurcations. We reveal a compatibility constraint between different out-of-equilibrium fluxes, which is a novel organizational principle, and explain how bifurcations are induced by turbulent fluctuations. 
Multistability and bifurcations of the mean flow are observed also in three dimensional turbulence~\cite{multistability_ravelet_2004,huisman_multiple_2014}, our work provides a clean example where the influence of turbulence on the stability of the mean flow can be understood.

\textbf{Setting} The LQG equation describes the strong rotation limit of a shallow fluid layer with a deformable surface, considering horizontal scales much larger than the deformation radius \cite{vallis_atmospheric_2017,larichev_weakly_1991,svirsky_statistics_2023}:
\begin{equation}
\label{eq:LQG}
\partial_\tau\psi + \bm{v}^\omega \cdot \bm{\nabla} \psi=\partial_{\tau}\psi+J(\omega,\psi)=f +\alpha\nabla^{2}\psi-\nu ( -\nabla^{2})^{p}\psi,
\end{equation}
where $J(\omega,\psi)=\partial_{x}\omega\partial_{y}\psi-\partial_{y}\omega\partial_{x}\psi$, $\psi$ is the stream-function and is proportional to surface perturbations of the layer, $\omega = \nabla^2 \psi$ is the vorticity, $\bm{v}^\omega = \bm{\hat{z} \times \bm{\nabla}}\omega$ is an effective velocity, $f$ is a forcing term, and the last two terms are dissipative --- a friction term due to linear drag acting on the velocity, and hyper-viscosity. The former provides the dominant dissipation mechanism at large scales, while the latter will dominate at small scales. In this system, the two inviscid quadratic invariants which lead to the formation of a condensate are the kinetic energy $Z=\frac{1}{2}\int\left(\nabla\psi\right)^{2}\text{d}^{2}x$, transferred from large to small scales (direct cascade), and the potential energy $E=\frac{1}{2}\int\psi^{2}\text{d}^{2}x$, transferred from small to large scales~\cite{smith_turbulent_2002}.
Note that the left-hand-side of \eqref{eq:LQG} arises as a limit of the Charney-Hasegawa-Mima equation, also describing magnetized plasmas and electron MHD~\cite{Hasegawa_pseudo_1978,Diamond_vorticity_2011}. In the former context Eq.~\eqref{eq:LQG} describes dynamics at scales much larger than the ion gyro radius, while in the latter, scales larger than the electron skin depth. 

In \cite{svirsky_statistics_2023,svirsky_two-dimensional_2023} we studied \eqref{eq:LQG} in the condensation regime, where the potential energy accumulates at the box scale $L$, serving as the IR cutoff. There is then a constant flux of potential energy in the range of scales $l_f>l>L$, termed here the inverse inertial range, where $l_f$ is the forcing scale. A constant flux of kinetic energy occurs in the range $l_f>l>l_\nu$, termed the direct inertial range, where $l_\nu$ is the UV cutoff of the inviscid system. At $l_\nu$ the energy transfer and the viscous dissipation rates are comparable, resulting in $l_{\nu} \sim (\nu^3/\eta)^{1/(6p-8)}$, where $\eta = \av{\bm{\nabla}\psi \cdot \bm{\nabla}f}$ is the kinetic energy injection rate, and $p$ is the hyper-viscous parameter. Note that the $\eta$ dose not depend on the large-scale flow configuration because a white in time forcing is used, hance $l_\nu$ can be viewed as an external control parameter.
 
We perform direct numerical simulations (DNS) integrating Eq.~\eqref{eq:LQG} using the Dedalus framework \cite{burns_dedalus_2020}, as described in \cite{svirsky_statistics_2023} but in a square domain $L\equiv L_{y}=L_x=2\pi$. Here we fix $\alpha=10^{-3}$ and $p=7$, while $l_f$ and $\nu$ are varied. We work in the regime ensuring turbulence and condensation~\cite{svirsky_statistics_2023}:  $\delta \equiv  \alpha(L^2/\epsilon)^{1/3} \ll1$, where $\epsilon = \langle \psi f \rangle$ is the potential energy injection rate, and $\text{Re}  \equiv  l_f^{2p-8/3}\epsilon^{1/3}/\nu \gg 1$. Statistics are gathered over many large-scale turnover times in a statistically steady state, defined by $(\tau_\alpha/E)(dE/dt) <  1$, where $\tau_{\alpha} = \alpha^{-1}L^2$ is the longest timescale in the system. Further details are given in \cite{SI}.

 \begin{figure}[t!]
    \centering
    \includegraphics[width=1\columnwidth]{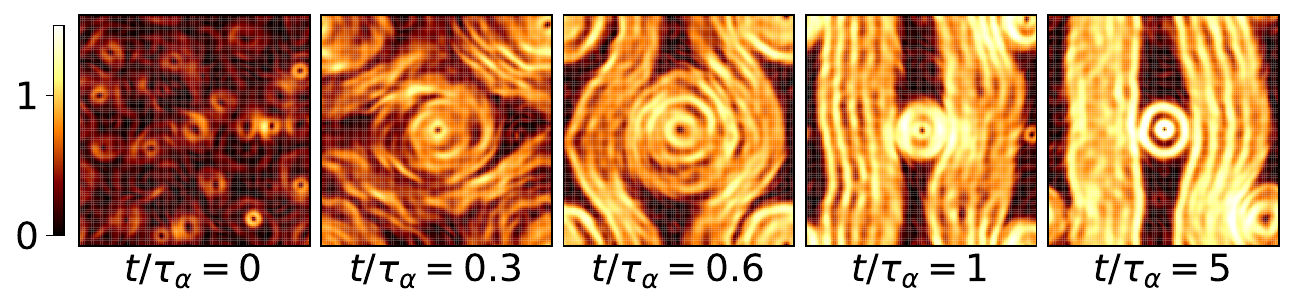}
    \caption{
        \label{fig:sym_breaking} 
        The formation of a jet condensate in a square domain in LQG. Snapshots of the normalized velocity magnitude $|\hat{z}\times\nabla \psi|/\sqrt{\epsilon/\alpha}$ are shown at different times .
        }
\end{figure}
Before discussing the results of simulations, we outline the anticipated outcomes. The condensate is a mean flow, characterized by $\Psi(\bm{x}) = \av{\psi}\neq 0$ --- the time-averaged stream-function. When the condensate is strong ($\delta\ll1$), and assuming it is stationary, it is constrained at leading order to be a stationary solution of the Euler equation, $J(\nabla^2\Psi,\Psi) = 0$, as in 2DNS~\cite{frishman_culmination_2017}.
\begin{figure}[h! t!]
    \centering
    \includegraphics[width=1\columnwidth]{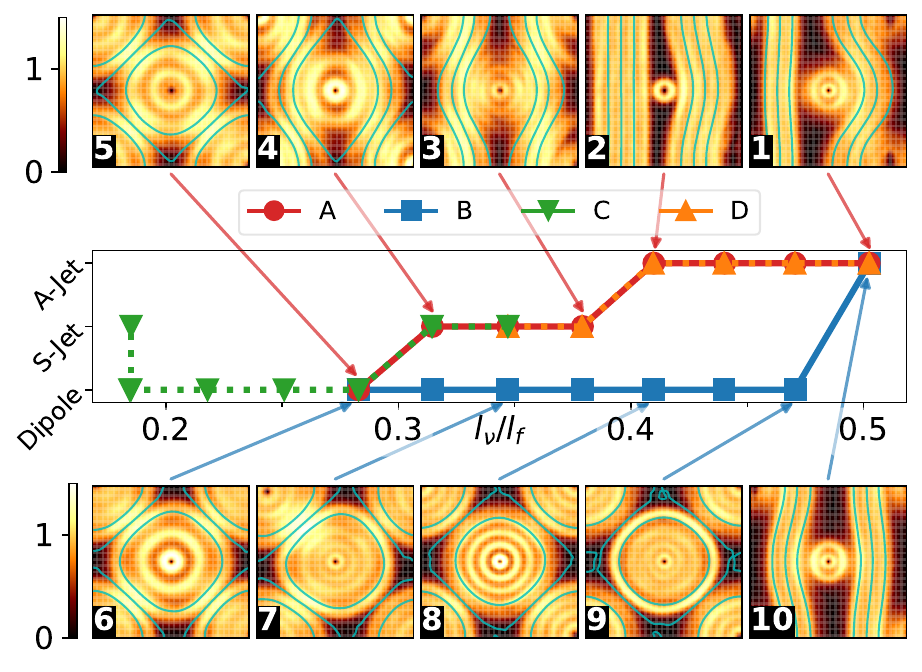}
    \caption{\label{fig:part1_scan_nu=1e-17} Classification of steady state condensate configurations into dipole (subplots 5, 6-9), symmetric jets (S-Jet, subplots 3,4) and anti-symmetric jets (A-Jet, subplots 1,2,10). Obtained by slowly varying $l_f$ while keeping $\nu = 10^{-17}$ fixed, the steps are indicated by colored symbols. Trajectory A is initiated from an A-Jet and $l_\nu/l_f$ is decreased, trajectory B is initiated from a dipole and $l_\nu/l_f$ is increased, and trajectories C/D are started from an S-Jet and $l_\nu/l_f$ is decreased/increased. 
    At $l_\nu/l_f < 0.2$ the condensate rapidly switches between dipole and S-Jet, indicated by marking both states. 
    Pictures at the top and bottom rows show the velocity magnitude $U/\sqrt{\epsilon/\alpha}$ and three streamlines
    at different locations of the trajectories $A$ (top) and $B$ (bottom), averaged over the eddy-turnover time at $l_f$.
    }
\end{figure}
In the absence of explicit symmetry breaking in the equations, the aspect ratio $L_y/L_x$ of the periodic domain is the control parameter determining the condensate configuration. In 2DNS two large-scale vortices, termed a dipole, are observed for $L_x=L_y$, two jets for $L_y > L_x$, and co-existence of the two for $1>L_y/L_x \gtrsim 1.1$ \cite{bouchet_random_2009,xu_fluctuation-induced_2023}. Similarly, in LQG for $L_y/L_x = 2$ we previously found jets~\cite{svirsky_statistics_2023,svirsky_two-dimensional_2023}.
In LQG, the magnitude of the condensate velocity $U = |\hat{z} \times \bm{\nabla} \Psi|$ can be determined using perturbation theory, even without reference to the geometry of the selected solution~\cite{SI}. 
This gives the prediction $U(\bm{x})= \sqrt{\epsilon/\alpha}$ as an exact result, and we expect a vortex dipole in our square domain.     

\textbf{Mean-flow configurations} Initiating simulations from zero initial conditions (IC), we find that two large vortices indeed always start forming, Fig.~\ref{fig:sym_breaking}. However, for some values of $(l_f,l_\nu)$, two jets later emerge between the vortices, their direction randomly selected during the build-up. The resulting mean flow breaks the $\pi/2$ rotational symmetry of the domain. Similar symmetry breaking had been previously observed  in decaying $\alpha$-turbulence~\cite{venaille_violent_2015}.

To explore the possible condensate configurations, we begin with this steady state as an initial condition and vary $l_f$ adiabatically while keeping $\nu$ fixed, allowing the system to reach a steady state between consecutive shifts. This results in trajectory $A$ in Fig.~\ref{fig:part1_scan_nu=1e-17}. The resulting configurations are shown in Fig.~\ref{fig:part1_scan_nu=1e-17} subplots 1-5, and are classified according to their symmetries~\cite{SI}. Initially, the condensate remains in the \emph{Asymmetric Jets} (A-Jet) configuration: one straight and one bent jet, and two small vortices in-between (Fig.~\ref{fig:part1_scan_nu=1e-17} subplots 1-2). Decreasing $l_\nu/l_f$ below $0.4$, the condensate now restores the $\pi$ rotational symmetry, taking the \emph{Symmetric Jets} (S-Jet) configuration with two symmetrically bending jets and larger vortices  (Fig.~\ref{fig:part1_scan_nu=1e-17} subplots 3-4). Finally, the $\pi/2$ rotational symmetry of the domain is restored for $l_\nu/l_f=0.325$, the condensate taking the form of a \emph{Dipole}  (Fig.~\ref{fig:part1_scan_nu=1e-17} subplot 5). 

Having arrived at a dipole configuration, we now check for hysteresis. We initiate a quasi-stationary trajectory from a dipole at $l_\nu/l_f=0.325$ and then gradually increase this parameter, trajectory $B$ in Fig.~\ref{fig:part1_scan_nu=1e-17} and subplots 6-10.
The condensate remains in the dipole configuration until it abruptly switches to the A-Jet at $l_\nu/l_f\approx0.5$ (configurations 9-10). Thus, there is coexistence of the dipole and the jets states in the range $0.325<l_\nu/l_f<0.475$, corresponding to a region of bi-stability. We have also initiated  trajectories from the S-Jet ($C$ and $D$) and found they coincide with the A-Jet (trajectory $A$), suggesting these states are connected by a continuous transition (configurations 2-3). Finally, we continued the dipole state below $l_\nu/l_f=0.325$, trajectory $C$, and found that for $l_\nu/l_f < 0.2$ it switches back and forth between a dipole and an S-Jet. We find this is a memoryless process~\cite{SI}, occurring at a rate faster than $\tau_\alpha$. We interpret this regime as one where the dipole is unstable: small ambient fluctuations are enough to induce a transition to a jet state, which is also unstable in this sense. We label such condensates as \emph{time-dependent}. 

\textbf{ Out-of-equilibrium constraint} To understand what determines the condensate configuration it is helpful to notice a key feature. In Fig.~\ref{fig:part1_scan_nu=1e-17} the velocity heatmaps in (1-10) are normalized according to our leading order prediction  $U/\sqrt{\epsilon/\alpha}=1$. The flow in most of the domain conforms with this, modulated by small-amplitude oscillations (as seen in Fig.~\ref{fig:sym_breaking} and discussed in \cite{svirsky_two-dimensional_2023}). However, black regions where $U/\sqrt{\epsilon/\alpha}\ll1$ are effectively devoid of a condensate, and occupy an increasingly larger fraction of the domain as $l_\nu/l_f$ increases. Indeed, for the configurations in (5-2) the vortex at the center continuously shrinks and is replaced by empty areas. It reaches its minimal size (set by $l_f/L$) in configuration 2, and beyond this point the vortex starts increasing at the expense of the jets, leaving behind even larger condensate voids. Similarly, for the dipole configurations (6-9) the area of the vortices is seen to systematically decrease as $l_\nu/l_f$ is increased. These observations suggest that the chosen configuration is determined by the area fraction of condensate voids, $\bar{C}$, which in turn is determined by $l_\nu/l_f$. 

\begin{figure}[ t!]
    \centering
    \includegraphics[width=1\columnwidth]{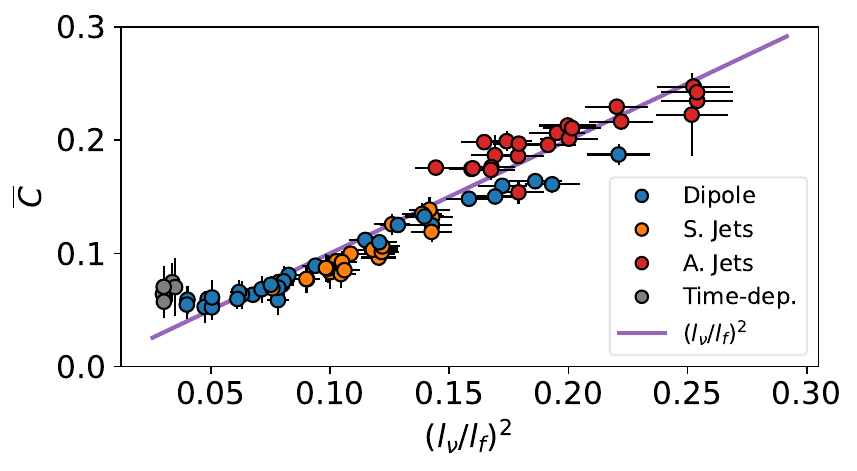}
    \caption{\label{fig:Area_vs_lnu_lf} The measured domain fraction of condensate voids, testing Eq.~\eqref{eq:area_fraction}. The colors indicate the observed steady state configuration of the mean flow. }
\end{figure}

We now show this is indeed the case.
In our previous work on LQG \cite{svirsky_statistics_2023} we found strong spatial fluxes of kinetic energy away from strong condensate areas. This remains true for all configurations we encountered, and can be traced to the local-in-space structure of the LQG dynamics, with the direction of the kinetic energy flux inherited from the direction of potential energy transfer in the condensate~\cite{SI}. We expect a similar effect to occur in QG with finite $L_d$ whenever the energy injection is such that  $l_f\gg L_d$. An important consequence of this spatial flux is that the kinetic energy cascade to small scales (and the small-scale dissipation) is then limited to the condensate voids, whose area fraction is $\bar{C}$. The latter can then be determined by a matching between the potential energy dissipation and input rates in it. 
The potential energy dissipation in condensate voids, denoted by $D_\nu^E$, occurs at small scales. To estimate it we relate it to the dissipation rate of the kinetic energy per unit area, $D_\nu^Z$: $D_\nu^Z = l_\nu^2 D_\nu^E$, assuming both occur at the scale $l_\nu$. Now, the injected kinetic energy in the entire domain gets dissipated in $\bar{C}$ so: $\eta =D_\nu^Z \bar{C}$. Taking into account that the forcing has a typical scale, so that $\epsilon=l_f^2 \eta$, we finally get $\bar{C} D_\nu^E=(l_\nu^2/l_f^2)\epsilon$ for the total (fraction) of potential energy dissipated in the voids per unit time.
We now equate this with the total injected potential energy in the voids per unit time, per unit area: $\epsilon \bar{C}$ (there are no spatial fluxes of potential energy from/to regions where the condensate is strong), and get the prediction:
\begin{equation}
    \label{eq:area_fraction}
    \bar{C} \approx (l_\nu / l_f)^2 ,
\end{equation}
We measure $\bar{C}$ in the different simulation runs, defined using a threshold on the normalized velocity magnitude $U$ 
as detailed in \cite{SI}. We find the results to be in good agreement with Eq.~\ref{eq:area_fraction} (Fig.~\ref{fig:Area_vs_lnu_lf}). 

\begin{figure*}[h! t!]
    \centering
    \includegraphics[width=2\columnwidth]{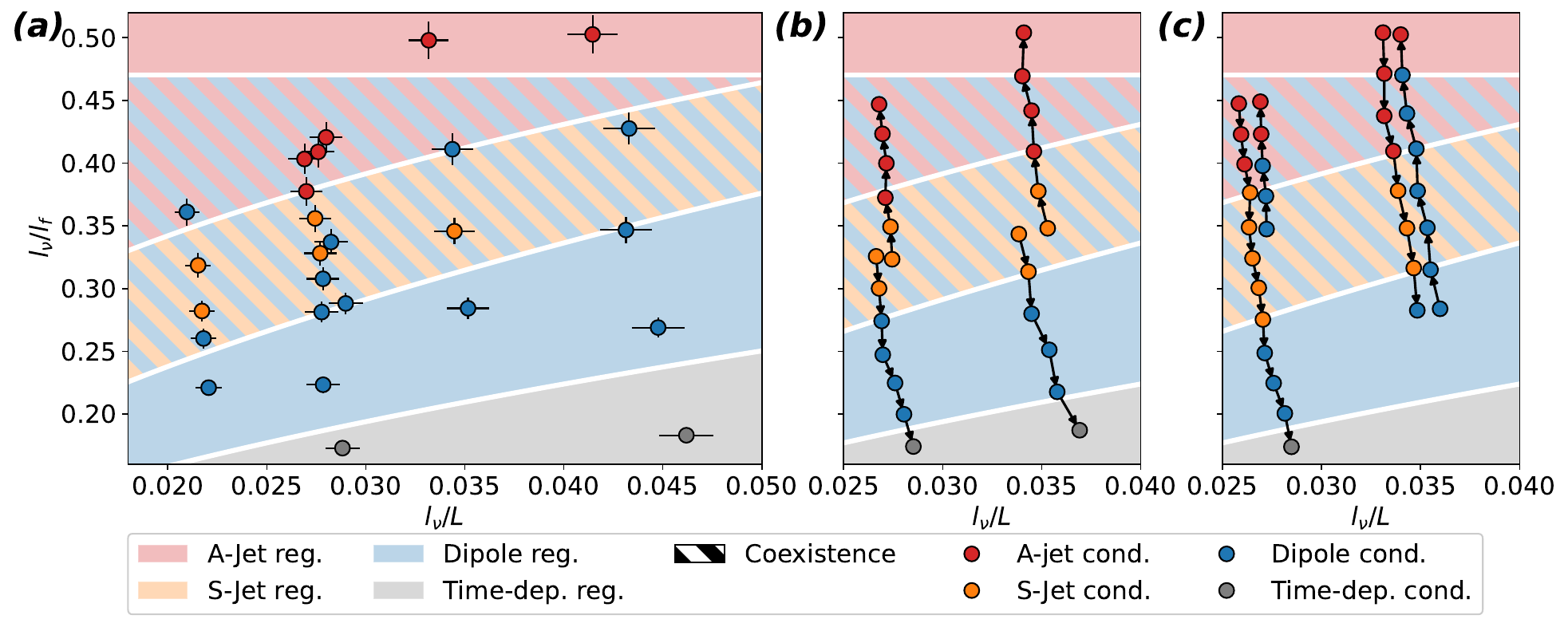}
    \caption{
        Phase diagram of condensate configurations, plotted in $(l_
        \nu/l_f, l_
        \nu/L)$ variables for clarity. The points correspond to steady state simulations, the color indicating the observed configuration. The colored regions mark theoretical predictions for existence/stability of the solutions. (a) Simulations initiated with $\psi = 0$. (b-c) Adiabatic trajectories, the steady-state of the simulation at the base of each arrow was used as the IC for the one at its tip. Simulations starting from the S-Jet and  A-Jet/dipole are shown in (b) and (c) respectively.   
        }
    \label{fig:phase_diagram}
\end{figure*}

\textbf{Stability of configurations} We now discuss what determines the stable configurations --- configurations which are persistent in time, i.e. realizable steady states. The dipole (jet) will lose its stability when typical fluctuations (always present as the system is stochastic and turbulent) are able to open (close) the streamlines, thus inducing a transition to the jet (dipole) state. The places where such topological changes can occur are the stagnation points (at the boundaries between vortices for both configurations), which are thus located in a condensate void. This turns the question of stability of a configuration into a geometrical one: there is a minimal size of the void below which fluctuations can bridge the gap and open (or close) the streamlines. Qualitatively, we expect transitions to be possible once the void area fraction $\bar{C}<\bar{C}_{\text{min}}\sim l_f^2/L^2$ since typical fluctuations are at the forcing scale. Combined with the expression for $\bar{C}$ in Eq.~\eqref{eq:area_fraction}, this implies states should tend to lose stability with decreasing $l_\nu/l_f$.  

\textbf{Phase diagram} The phase diagram of states is two dimensional, determined by the extent of the direct inertial range $l_\nu/l_f$, and that of the inverse inertial range $l_f/L$. We use the equivalent variables $(l_\nu / l_f,l_\nu/L)$ for better visualization in Fig.~\ref{fig:phase_diagram} where we present it.
We estimate the minimal void areas empirically: jets become unstable when $\bar{C}=l_\nu^2/l_f^2 < 8l_f^2/L^2$ (corresponding to two rectangles of area $l_f\times 4l_f$, Fig.~\ref{fig:part1_scan_nu=1e-17} subplot $5$), while dipoles become unstable when $\bar{C}=l_\nu^2/l_f^2<2 \pi (l_f/L)^2$ (corresponding to two discs of diameter $l_f$), replaced by the time-dependent condensate state. This gives the lower part of the phase diagram in Fig.~\ref{fig:phase_diagram}, including the two lowest boundaries. The third boundary, between the S-Jet (coexisting with the dipole) and the A-Jet (coexisting with the dipole) occurs when the vortex between the jets takes its smallest size ($\propto l_f^2$), Fig.~\ref{fig:part1_scan_nu=1e-17} subplot 2. This boundary is found by estimating the area fraction of the void in this configuration as $\bar{C}=l_\nu^2/l_f^2=2l_f / L$ (two rectangles of area $l_f\times L$).  
Condensate configurations in simulations initiated from $\psi=0$ IC (Fig.~\ref{fig:phase_diagram}(a)) indeed reside in the regions allowed by the above arguments. Moreover, overlaying adiabatic trajectories in parameter space (Fig.~\ref{fig:phase_diagram}(b-c)) demonstrates that the derived boundaries are precisely where the condensate switches between configurations.    

Going further up in the phase diagram: increasing $l_\nu/l_f$ increases the void area fraction, and the dipole state eventually ceases to exist. This is a consequence of principle (i) for the condensate: that it should take the largest available scale.  
For the dipole configuration, the minimal possible condensate area that still spans the entire domain is that of two tightly packed vortices (Fig.~\ref{fig:part1_scan_nu=1e-17} subplot 9), giving a lower bound for the condensate area fraction: $1-\bar{C}=\pi/4$. For $\bar{C}=(l_\nu/l_f)^2>1-\pi/4$, hence only the A-Jet configuration remains, giving the upper most boundary in Fig.~\ref{fig:phase_diagram}. Our simulations are consistent with these predictions except for one run (increasing $l_\nu / l_f$, at $l_\nu \approx 0.027$ in Fig.~\ref{fig:phase_diagram}(c)), which switches from the dipole to the A-Jet earlier than predicted, but still in the allowed region for A-Jets. Note that $l_\nu/l_f$ cannot be arbitrarily increased and we are always assuming a sufficient scale separation ($Re=(l_f/l_\nu)^{(6p-8)/3}\gg 1$).

Considering the above boundaries in the "thermodynamic" infinite domain limit
$L\gg l_f$: if $(l_\nu/l_f)^2\ll l_f/L$ (the direct inertial range longer than the inverse one) the only remaining configuration is the time-dependent condensate. On the other hand, if $(l_\nu/l_f)^2\gg l_f/L$, two phases remain: A-Jets for $l_\nu/l_f \gtrsim
0.4$ (coexistence of four types: horizontal/vertical jets, left/right bent jet) and a coexistence phase between A-Jets and a dipole for $l_\nu/l_f \lesssim 0.4 $. 

Then, for $l_\nu/l_f \gtrsim
0.4$ where only jets exist, any particular state breaks the $\pi/2$ symmetry of the domain. This symmetry could be statistically restored if the system randomly switches between states, but we have not observed such transitions for A-Jets. Instead, we suggest that the systems' symmetry is spontaneously broken. Indeed, a transition from e.g. horizontal jets to vertical ones requires closing the condensate voids and reopening them in the other direction. However, in the A-Jet state the length of the condensate voids parallel to the jets is extensive ($\propto L$), while the size of typical fluctuations is $l_f$. Therefore, closing the voids requires a fluctuation whose probability is expected to be exponentially suppressed as $l_f/L\to 0$, implying spontaneous symmetry breaking in the thermodynamic limit. 

\textbf{Conclusion} We found an unexpectedly rich landscape of large-scale flow configurations, absent in 2D Navier-Stokes. Its existence can be traced to the enslavement of the small-scale (direct) flux to the large-scale (inverse) flux occurring in LQG~\cite{svirsky_two-dimensional_2023}. We expect similar effects for flows in the regime $\rm{Fr}^2\ll\rm{Ro}\ll \rm{Fr}\ll1$ when $l_f>L_d$, as the dynamics on scales larger than $L_d$ then follow LQG~\cite{polvani_coherent_1994,smith_turbulent_2002}. The range of interactions is thus a relevant control parameter shaping coherent structures; when it is small enough, the interplay between out-of-equilibrium fluxes imposes a novel organizational principle.

\begin{acknowledgments}
\section{Acknowledgments}
This work was supported by BSF grant No. 2022107 and ISF grant No. 486/23.
\end{acknowledgments}

\bibliography{ref}

\end{document}